\documentclass[a4paper,fleqn]{cas-dc}

\usepackage{amsmath}
\usepackage[numbers,square,sort&compress]{natbib}
\usepackage[normalem]{ulem}
\usepackage{hyperref}
\def\tsc#1{\csdef{#1}{\textsc{\lowercase{#1}}\xspace}}
\tsc{WGM}
\tsc{QE}

\begin{document}
%%% CUSTOM COMMANDS %%%
\newcommand{\bmat}[1]{\mathrm{\textbf{#1}}}
\newcommand{\bvec}[1]{\textit{\textbf{#1}}}
\newcommand{\de}{\mathrm{d}}
\newcommand{\nptm}{\texttt{NP\_TMcode}}
%%% END OF CUSTOM COMMANDS %%%

\let\WriteBookmarks\relax
\def\floatpagepagefraction{1}
\def\textpagefraction{.001}

\shorttitle{Nano-particle T-matrix for extinction in ISM and exoplanetary atmospheres}

\shortauthors{G. La Mura et al.}

\title [mode = title]{Extinction by non-spherical particles with the Nanoparticle Transition Matrix Code: perspectives for interstellar medium and exoplanetary atmospheres}

\author[1]{{G.} {La Mura}}[orcid=0000-0001-8553-499X]

\cormark[1]

\ead{giovanni.lamura@inaf.it}

\credit{Software development, manuscript writing}

\cortext[1]{Corresponding author}

\author[1]{G. Mulas}[orcid=0000-0003-0602-6669]

\ead{giacomo.mulas@inaf.it}

\credit{software optimization, conceptualization of the study}

\author[2]{M. A. Iat\`{i}}[orcid=0000-0002-3576-8656]

\ead{mariaantonia.iati@cnr.it}

\credit{model development, physical interpretation}

\author[3]{C. Cecchi-Pestellini}[orcid=0000-0001-7480-0324]

\ead{cesare.cecchipestellini@inaf.it}

\credit{dielectric function calculations}

\author[1]{G. Aresu}[orcid=0000-0003-3683-0834]

\ead{giambattista.aresu@inaf.it}

\credit{model calculations}

\author[2]{S. Rezaei}[orcid=0009-0007-0916-9469]

\ead{shadi.rezaei@unime.it}

\credit{performance evaluation}

\author[4]{R. Saija}[orcid=0000-0002-5823-9749]

\ead{rosalba.saija@unime.it}

\credit{theoretical framework, physical interpretation}

\affiliation[1]{organization={INAF - Astronomical Observatory of Cagliari},
            addressline={Via della Scienza 5}, 
            city={Selargius},
            postcode={09047}, 
            state={CA},
            country={Italy}}

\affiliation[2]{organization={CNR-IPCF, Institute for Chemical and Physical Processes},
            addressline={Viale Stagno D'Alcontres 37}, 
            city={Messina},
            postcode={98158}, 
            state={ME},
            country={Italy}}

\affiliation[3]{organization={INAF - Astronomical Observatory of Palermo},
            addressline={Piazza del Parlamento 1}, 
            city={Palermo},
            postcode={90134}, 
            state={PA},
            country={Italy}}

\affiliation[4]{organization={Department of Mathematics and Computer Sciences, Physical Sciences and Earth Sciences - University of Messina},
            addressline={Viale Stagno D'Alcontres 31}, 
            city={Messina},
            postcode={98166}, 
            state={ME},
            country={Italy}}

\begin{abstract}
The multi-sphere Transition Matrix formalism (T-matrix) offers an optimal solution to the problem of radiation scattering and absorption by particles characterized by a high degree of chemical and structural complexity. However, its extensive application to particle models composed by large numbers of constituents has been classically limited by the substantial computational burden of the method. As a consequence, the investigation of light scattering and absorption processes, in cases where particle asymmetry and non-homogeneity play an important role, typically need to rely on simplifying assumptions which affect the correct interpretation of transmission spectra and extinction curves. This problem has a huge impact on many fields of scientific application. In particular, it has a critical role in astrophysics, because the ability to model highly non homogeneous particles is a fundamental requirement to understand the role of extinction in different phases of the interstellar medium (ISM), as well as to probe the physics of light transmission in planetary atmospheres within the Solar System and beyond.

In this work we describe the application of a new implementation of the T-matrix formalism, based on the framework originally developed by Borghese, Denti \& Saija and named \textit{Nano-particle Transition Matrix code}, which, taking advantage of parallel and scalable algorithms, leverages the power of modern hardware architectures to reduce the computational costs related to the solution of examples of relevant particle models for the investigation of dust in the ISM and of hot-Jupyter class planets atmospheric aerosols. This new implementation of the T-matrix formalism can easily handle calculations of increasing complexity with a direct impact in the interpretation of advanced spectroscopic observations, such as the transmission spectra of transiting exoplanets produced by the \textit{James Webb Space Telescope} (JWST) and those expected from the \textit{Atmospheric Remote-sensing Infrared Exoplanet Large survey} (ARIEL).
\end{abstract}

\begin{graphicalabstract}
\includegraphics[width=\textwidth]{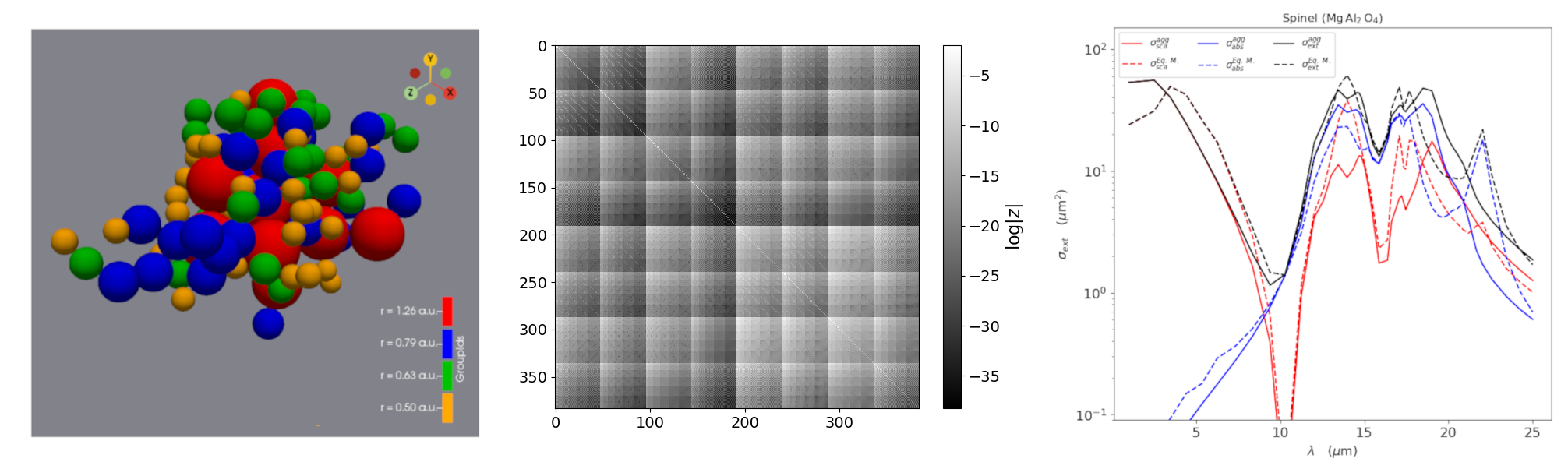}
\end{graphicalabstract}

\begin{highlights}
\item New parallel implementation of the T-matrix formalism.
\item Light scattering and absorption.
\item Modeling extinction properties of complex dust particles in astrophysical environments.
\item Modeling particles in exo-planetary atmospheres.
\end{highlights}


\maketitle

\section{Introduction}
\label{sec-intro}
The study of electromagnetic radiation propagation through non-homogeneous transmissive media has grown to acquire fundamental importance for many fields of modern scientific and technological research. While homogeneous approximations suffice for large-scale descriptions, the presence of localized inhomogeneities - such as material particles with diverse morphologies and compositions - enforces a rigorous, detailed treatment to correctly interpret observational data. This problem is particularly relevant to astrophysics, because the extinction process, resulting from radiation scattering and absorption by interstellar dust, affects our understanding of the cosmic environment. In recent times, the exploration of exoplanetary atmospheres via transmission spectroscopy has further highlighted the importance of this field. In this context, suspended aerosols play a decisive role in modulating atmospheric transmittance across various wavelengths \citep{Gao21}, making the accurate determination of absorption and scattering efficiencies essential for connecting the observed spectral signatures to the underlying chemical compositions of the observed targets.

The typical effects of radiation-matter interactions are generally described in terms of the ratio between the characteristic particle dimension $d$ and the radiation wavelength $\lambda$. While the limits of geometric optics (valid when $\lambda \ll d$) and the Rayleigh regime (applying when $\lambda \gg d$, with $\sigma_{sca} \propto \lambda^{-4}$) offer simplified analytical solutions, the intermediate regime where $\lambda \sim d$ requires a complete electromagnetic treatment through the solution of Maxwell's equations. The theoretical grounds for the intermediate size regime were formulated within the \textit{Mie theory} \citep{Mie1908}, which provides an exact analytical solution for perfectly spherical particles. However, real-world scatterers like dust grains, soot, and complex aerosol droplets, often deviate significantly from spherical symmetry, making simple Mie-based models or basic geometric generalizations (such as cylinders or ellipsoids) inadequate for reproducing observed optical properties \citep{Lodge24}. To address these complexities, several numerical frameworks have been developed, leveraging the linearity of the radiation field equations. Among the main ones, we may recall the \textit{Discrete Dipole Approximation} \citep[DDA,][]{Purcell73,Draine94}, which divides the scatterer into a finite array of polarizable points, the \textit{Mean Field Theory} (MFT) and its subsequent evolutions \citep{Berry86, Botet97, Tazaki18}, particularly effective for fractal aggregates, the \textit{Finite Difference Time Domain} (FDTD) method \citep{Taflove05, Yee96}, which relies on the direct numerical integration of Maxwell's equations, and the \textit{Transition Matrix} (T-matrix) formalism \citep{Waterman71, Mishchenko96, Borghese07}, which stands out for its ability to model arbitrary clusters of spherically symmetric components with high precision.

\begin{figure*}
  \begin{center}
    \includegraphics[width=0.98\textwidth]{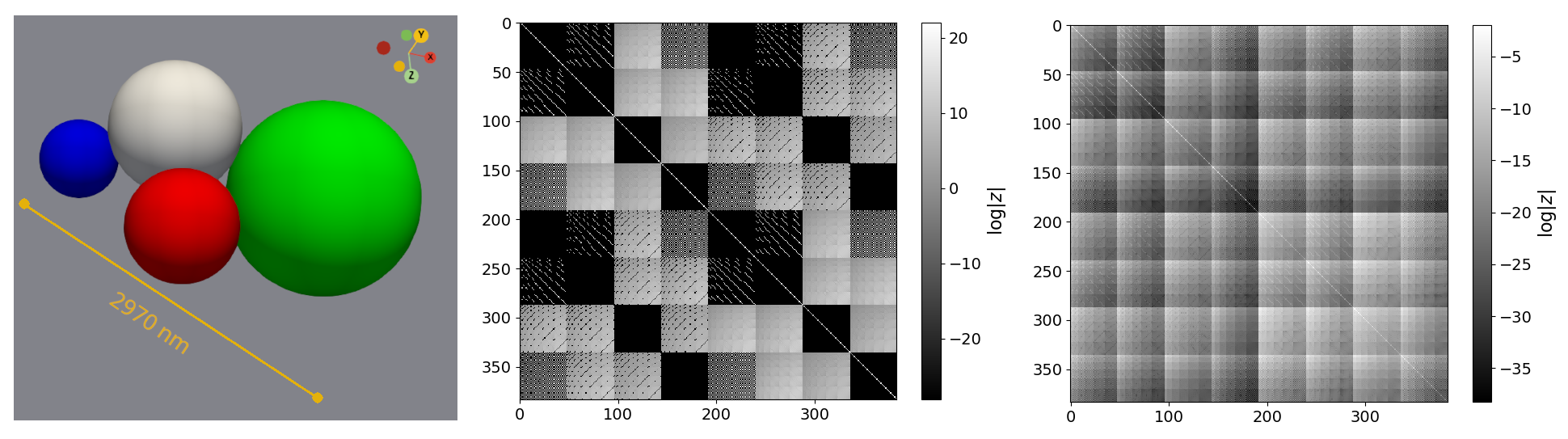}
  \end{center}
  \caption{Example of T-matrix representation for a simplified model particle. \textit{Left panel}: %\sout{a generic}
  3D rendering of an example particle, represented as an aggregate of $4$ spheres, each one marked by a different color. The size of the particle is shown by the labeled ruler. \textit{Central panel}: Gray-scale representation of the logarithm of the absolute values of the elements $z$ of the (not yet inverted) continuity matrix for a radiation field with wavelength $\lambda = 1.43\, \mu$m. \textit{Right panel}: Gray-scale representation of the logarithm of the absolute values of the elements $z$ of the multi-centered T-matrix. Dark areas correspond to small values. The T-matrix multi-polar expansion is truncated at order $6$. The matrix elements $z$ are complex numbers represented in double precision.\label{fig_dev_model}}
\end{figure*}
Although the T-matrix approach is particularly advantageous when dealing with structural complexity and with the requirement for orientation-averaged quantities, the derivation of the T-matrix is intrinsically characterized by a significant computational burden, because modeling an asymmetric particle requires taking into account a large number of spherical constituents and of their mutual interactions. As a consequence, the T-matrix must be obtained from the construction and the subsequent inversion of a possibly large continuity matrix (with thousands of lines and columns), requiring up to some gigabytes of memory only to store the matrix for every radiation wavelength. Nevertheless, the recent paradigm shift toward high-performance computing (HPC) and the widespread availability of parallel hardware architectures are now starting to provide a promising pathway to optimize this type of intensive calculations. In this work, we discuss the possible application of a revised implementation of the T-matrix formalism, based on the original theoretical framework of Borghese, Denti \& Saija \citep{Borghese07}, which we name \textit{Nano-Particle Transition Matrix code} \citep[\nptm,][]{LaMura25,LaMura26},\footnote{\nptm\ is distributed under the terms of the GNU GPLv3 license at \href{https://github.com/ICSC-Spoke3/NP_TMcode}{https://github.com/ICSC-Spoke3/NP\_TMcode}. The version used in this work is \texttt{NP\_TMcode~v0.10.10}.} to obtain extinction curves for a set of illustrative particle models, designed to represent typical structures expected in the interstellar medium (ISM) and in the atmospheres of hot-Jupyter exoplanets. Our work is organized as follows: in Section~\ref{sec-implementation}, we describe the software implementation of the T-matrix formalism; in Section~\ref{sec-applications}, we illustrate the application of the code to models of ISM dust grains and exoplanetary atmospheric aerosol particles; in Section~\ref{sec-perf}, we discuss the performance of the code and its requirements; finally, in Section~\ref{sec-discussion}, we provide our concluding remarks.

\section{Implementation of the T-matrix formalism}
\label{sec-implementation}

\nptm\ is made up by four core applications written in \textit{C++}, named \texttt{np\_sphere}, \texttt{np\_cluster}, \texttt{np\_inclusion}, and \texttt{np\_trapping}, together with a set of auxiliary \textit{Python} scripts, designed to help users in preparing their particle models and inspecting the calculation results. The simplest application of the suite, \texttt{np\_sphere}, solves the scattering problem from a single spherical particle with the \textit{Mie theory} \citep{Mie1908}. Conversely, \texttt{np\_cluster} and \texttt{np\_inclusion} apply the T-matrix formalism described by \citet{LaMura26}, respectively, for the case of an arbitrary aggregate of non-intersecting spherical particles and for an aggregate of spherical particles embedded in an outer spherical envelope. Finally, \texttt{np\_trapping} computes the dynamical effects of a focused laser beam interacting with a model particle, evaluating the forces and torques that it would experience close to the beam focal point.

\begin{figure}
  \begin{center}
    \includegraphics[width=0.49\textwidth]{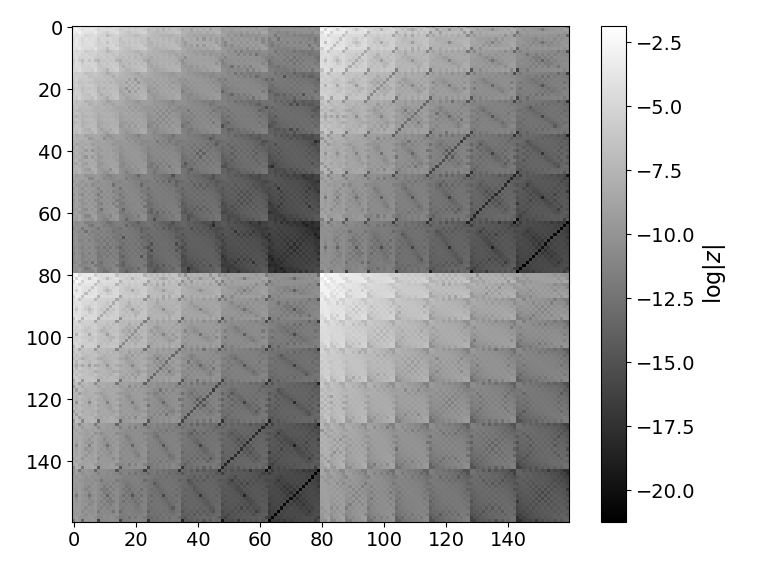}
  \end{center}
  \caption{Gray-scale representation of the logarithm of the absolute values of the elements of the single centered T-matrix of the example particle illustrated in Fig.~\ref{fig_dev_model} for a radiation field with wavelength $\lambda = 1.43\, \mu$m. Dark areas correspond to small values. The multi-polar expansion for the whole particle is truncated at order $8$. The matrix elements $z$ are complex numbers represented in double precision.\label{fig_dev_ztm}}
\end{figure}
Since the T-matrix formalism can use spherical monomers characterized by different size, composition and layer structure, it can approximate a wide variety of particle models. In its most fundamental implementation, the formalism represents a particle as a collection of spheres, whose scattering coefficients are used to build a multi-centered continuity matrix, which connects the incident fields with the corresponding scattered ones across the particle, by decomposing the fields on a set of multi-polar spherical vector harmonics. The inverse of this continuity matrix is the multi-centered operator that describes the scattering process from the point of view of the particle constituents. An example of this implementation is illustrated in Fig.~\ref{fig_dev_model}. Once the multi-centered T-matrix has been computed for a specific wavelength, the scattering amplitudes of the particle components can be recombined, thanks to the spherical vector harmonics addition theorem \citep{Borghese80}, to obtain the single centered T-matrix of the whole particle, which, for the case illustrated in Fig.~\ref{fig_dev_model}, yields the result shown in Fig.~\ref{fig_dev_ztm}. The transformation of the multi-centered T-matrix of the spherical components into the mono-centered model T-matrix approximates the T-matrix of a real particle with an accuracy that improves as far as the aggregate of spheres uses a large enough number of components to better reconstruct the particle's actual shape.

\begin{figure*}
  \begin{center}
    \includegraphics[width=0.95\textwidth]{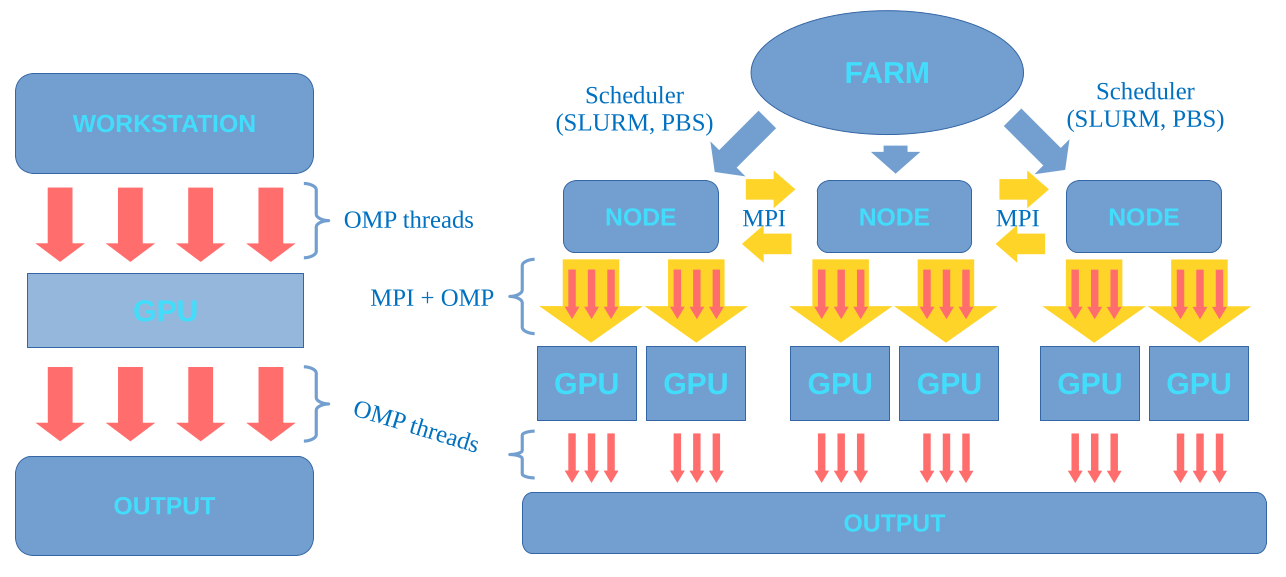}
  \end{center}
  \caption{Schematic representation of the parallelization strategies used by \nptm\ applications. Work-sharing tasks and operations that benefit from fast access to a common memory environment are managed via \texttt{OpenMP} threads (red arrows). Scalability across multiple nodes of a computing farm, instead, is obtained by distributing independent wavelength calculations through the \texttt{MPI} interface (yellow arrows), under the coordination of a job scheduler (blue arrows). If GPU devices are available, they can be optionally used to further speed up algebraic calculations, such as matrix inversion and matrix - vector algebra. \label{fig_parallel_scheme}}
\end{figure*}
From the analysis of the example case discussed in Fig.~\ref{fig_dev_model} and \ref{fig_dev_ztm}, it is clear that the complexity of the calculation mainly depends on four factors:
\begin{enumerate}
\item the number of spheres used to represent a particle (complex shapes are generally better represented by several small spheres);
\item the field truncation order for the interactions with the spherical components;
\item the field truncation order for the interaction with the whole particle;
\item the number of wavelengths for which the solution is needed.
\end{enumerate}
Given a particle model formed by $n_{sph}$ components, evaluated up to a maximum multi-polar expansion order $l_M$, the multi-centered continuity matrix (the entity which needs to be inverted) is a $[N \times N]$ complex matrix, with
\begin{equation}
  N = 2 n_{sph} l_M (l_M + 2). \label{eq_matrix_size}
\end{equation}
Since the terms entering the continuity matrix tend to grow in magnitude at large multipole orders and the required truncation order for an acceptable description of the particle $l_M$ increases with the particle's size parameter:
\begin{equation}
    x = \dfrac{2 \pi \sqrt{\varepsilon}}{\lambda}r, \label{eq_size_par}
\end{equation}
where $r$ denotes the particle radius, $\lambda$ the incident wavelength, and $\varepsilon$ the dielectric constant of the surrounding non-conductive medium, the matrix inversion step can soon become a computationally demanding task. Moreover, if the dynamic range of the elements of the matrix exceeds the precision limit of numeric types, such as double precision floating point numbers, the matrix inversion becomes numerically unstable. An estimate of the maximum expansion order $l_i$, needed to achieve convergence within the limits of double precision floating point numbers for the case of a single sphere, can be obtained following \citet{Wiscombe80}:
\begin{equation}
    l_i = x + 4.05\, x^{1/3} + 2. \label{eq_Wiscombe}
\end{equation}
This criterion is fairly robust and reliable for dielectric particles, though it may break down for highly conductive (i.~e. metallic) particles. For the case of an aggregate of spheres, Eq.~(\ref{eq_Wiscombe}) needs to be modified into:
\begin{equation}
  l_e = x + 11 x^{1/3} + 1, \label{eq_Wiscombe_ext}
\end{equation}
where, this time, $x$ must be computed using the radius of the smallest sphere which contains the particle $r_{agg}$, instead of $r$. Taking advantage of the fact that the particle constituents are by definition smaller than the whole particle, it has been shown that the description of the fields that develop within the particle can be obtained using a lower \textit{internal} order $l_i$, with respect to the \textit{external} order $l_e$, used for the entire particle \citep{Iati04}. In addition, if the solution is calculated for the same particle over a range of different wavelengths, where increasing values of $\lambda$ correspond to smaller values of $x$, it is in principle possible to further reduce the computational costs by dynamically evaluating the optimal values of $l_i$ and $l_e$ for every calculation step.

\begin{table*}
\caption{Summary of the geometrical properties of the models presented in this study. The table columns provide the name of the model, the number of spherical components used to build the particle, the radius of the smallest sphere that contains the aggregate ($r_{agg}$), the radius of the sphere having the same mass of the aggregate ($r_{eq.m.}$), the radii of all the types of different spheres used in the model ($r_1, ..., r_4$), and the scaling factors used to test various particle sizes. All radii and sphere positions were scaled according to the listed factors. \label{tab-models}}
\begin{tabular*}{\tblwidth}{@{}LCCCCCCCC@{}}
\toprule
 Model & $n_{sph}$ & $r_{agg}$ & $r_{eq.m.}$ & $r_1$ & $r_2$ & $r_3$ & $r_4$ & Scaling\\ % Table header row
\midrule
Naked Si-C & $46$ & $0.158\, \mu$m & $0.081\, \mu$m & $0.045\, \mu$m & $0.020\, \mu$m & -- & -- & $1 \times, 2 \times, 3 \times$ \\
Ice coated Si-C & $46$ & $0.158\, \mu$m & $0.158\, \mu$m & $0.045\, \mu$m & $0.020\, \mu$m & -- & -- & $1 \times, 2 \times, 3 \times$ \\
Exoplanetary & $100$ & $1.000\, \mu$m & $0.363\, \mu$m & $0.126\, \mu$m & $0.079\, \mu$m & $0.063\, \mu$m & $0.050\, \mu$m & $1 \times, 2 \times, 3 \times, 5 \times$  \\
\bottomrule
\end{tabular*}
\end{table*}
\nptm\ makes use of parallel computing architectures in order to significantly reduce the amount of time required to achieve the solution of high complexity particle models. The implementation strategy focuses on two main aspects, involving the optimization of work-sharing tasks and the simultaneous execution of independent calculations over different wavelengths, following the scheme illustrated in Fig.~\ref{fig_parallel_scheme}. The optimization of work-sharing tasks is obtained by linking the linear algebra calculations with highly specialized numeric libraries, such as \texttt{LAPACK} \citep{lapack99}, or MAGMA \citep[][used when a suitable GPU is available]{magma97}, as well as by concurrent execution of independent loop calculations via \texttt{OpenMP} multi-threading directives \citep{Chandra01}. The simultaneous execution of embarrassingly parallel, independent wavelength calculations, on the contrary, can either be managed by \texttt{OpenMP} hierarchical multi-threading, if the host system grants enough memory resources to handle more than one calculation at a time, or via multiple \texttt{MPI} processes \citep{MPI93}, if the calculation can be scattered across several computing nodes. At present, no multi-node solution of a single wavelength iteration is implemented, although this feature is currently under investigation for a possible implementation in future code releases.

In this study we present the extinction curves evaluated for a set of illustrative models designed to simulate the characteristic of interstellar dust particles and of hot-Jupyter planetary atmosphere condensates, exploring a range of different sizes. Our models are obtained by combining spherical components with different radii and compositions. Each model is characterized by the number of spheres $n_{sph}$, the radius of each type of sphere $r_i$ (which is used to calculate the optimal value of $l_i$), the radius of the smallest sphere that contains the aggregate $r_{agg}$ (which, instead, is used to calculate $l_e$), and the radius of the single sphere that contains the same amount of mass as the aggregate $r_{eq.m.}$ (which is used to compute the Mie solution). For multi-material particles, we adopted as equivalent case a sphere with multiple concentric layers, each containing the same amount of every material of the modeled particle and arranged to match as closely as possible the same sequence of materials in a path leading from the particle center to its surface. To further validate the results of our calculations, we constructed the same geometries of the smallest variants of our particle models and we computed their average extinction cross-sections, using the \textit{Amsterdam Discrete Dipole Approximation} code \citep[\texttt{ADDA~v1.5.0-alpha3},][]{Yurkin11}. A summary of the settings used to define our models is presented in Table~\ref{tab-models}. We repeated our calculations scaling the geometry of each model by different factors, to investigate the effects of particle size on radiation extinction.

\begin{figure*}
  \begin{center}
    \includegraphics[width=0.95\textwidth]{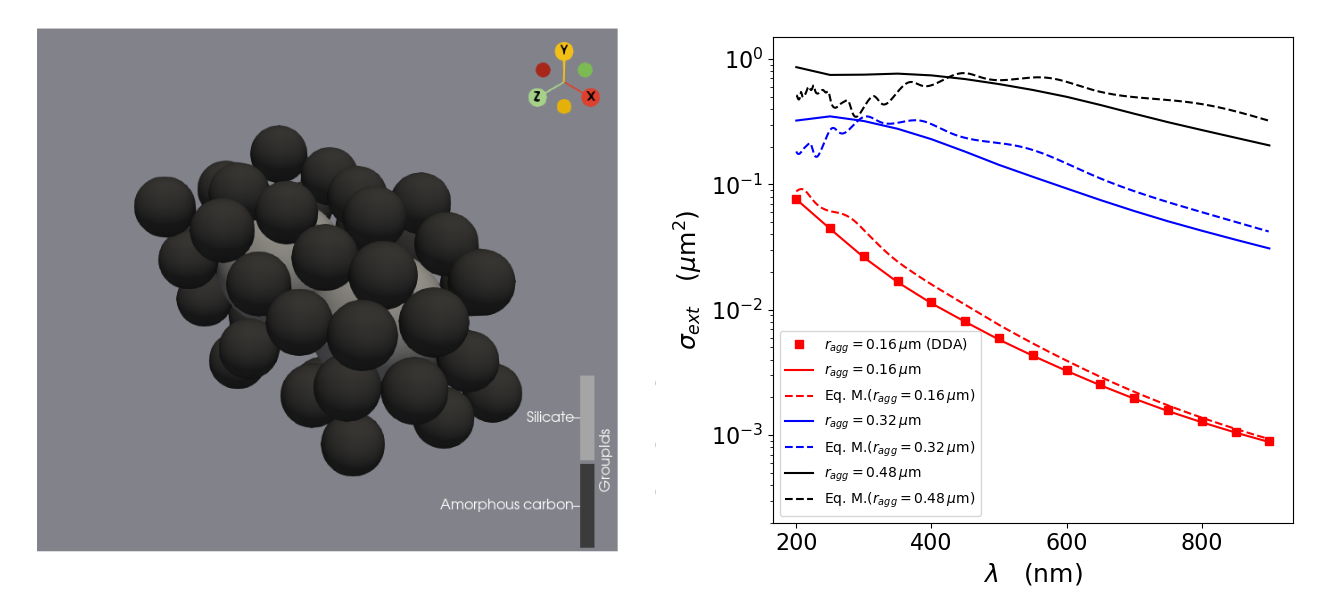}
  \end{center}
  \caption{\textit{Left panel:} 3D representation of an elongated ISM dust grain model composed by a silicate core made up by $2$ large spheres (here shown as inner gray spheres), surrounded by a layer of amorphous carbon, represented by $44$ smaller spheres (depicted as darker spheres, with $r_{Si} = 2.25 r_C$). The scale of the 3D model is arbitrary. \textit{Right panel:} Average extinction cross-sections, computed in the wavelength range $0.2\, \mu \mathrm{m} \leq \lambda \leq 0.9\, \mu \mathrm{m}$ for scaled versions of the model particle that fit within a cirular radius of $r_{agg} = 0.158\, \mu$m (red continuous line), $0.316\, \mu$m (blue continuous line), and $0.474\, \mu$m (black continuous line). All the particle models are compared with the corresponding Mie solutions for spherical particles of equivalent mass (dashed lines). The smallest particle is also compared with the average extinction cross-section computed with the \texttt{ADDA} implementation of the DDA formalism (red squares). Increasing deviations from the spherical case can be observed for larger particle sizes. \label{fig_mod02}}
\end{figure*}
\section{Applications}
\label{sec-applications}

\subsection{Interstellar dust grains}
\label{sec-dust}

Dust is an important ingredient of the ISM. In spite of accounting for approximately only one percent of the ISM mass in the Milky Way, it has a key role in many fundamental aspects of astrophysics. In addition to being the main cause of interstellar extinction in the optical and UV wavelength ranges, it represents the largest reservoir of refractory elements in the diffuse space, it participates to the thermal balance of the ISM, and it hosts chemical processes that are responsible for the formation of complex molecules in outer space environments. Although it is widely accepted that ISM dust grains are mainly sub-micrometer  particles, composed by silicate and carbonaceous materials \citep{Aiello00, Mulas13, Mishra15}, possibly combined with volatile ices in dense and cold environments, the details of their distribution are still matter of debate. The problem of whether the various classes of materials are segregated in different populations of grains \citep[e.g.][]{Weingartner01} or intermixed in the grain structure \citep{Li97, Hensley23} requires advanced modeling capabilities, since a change in the properties of dust grains is expected to leave observable signatures in the interstellar extinction process \citep{Cecchi-Pestellini14}.

\begin{figure*}
  \begin{center}
    \includegraphics[width=0.95\textwidth]{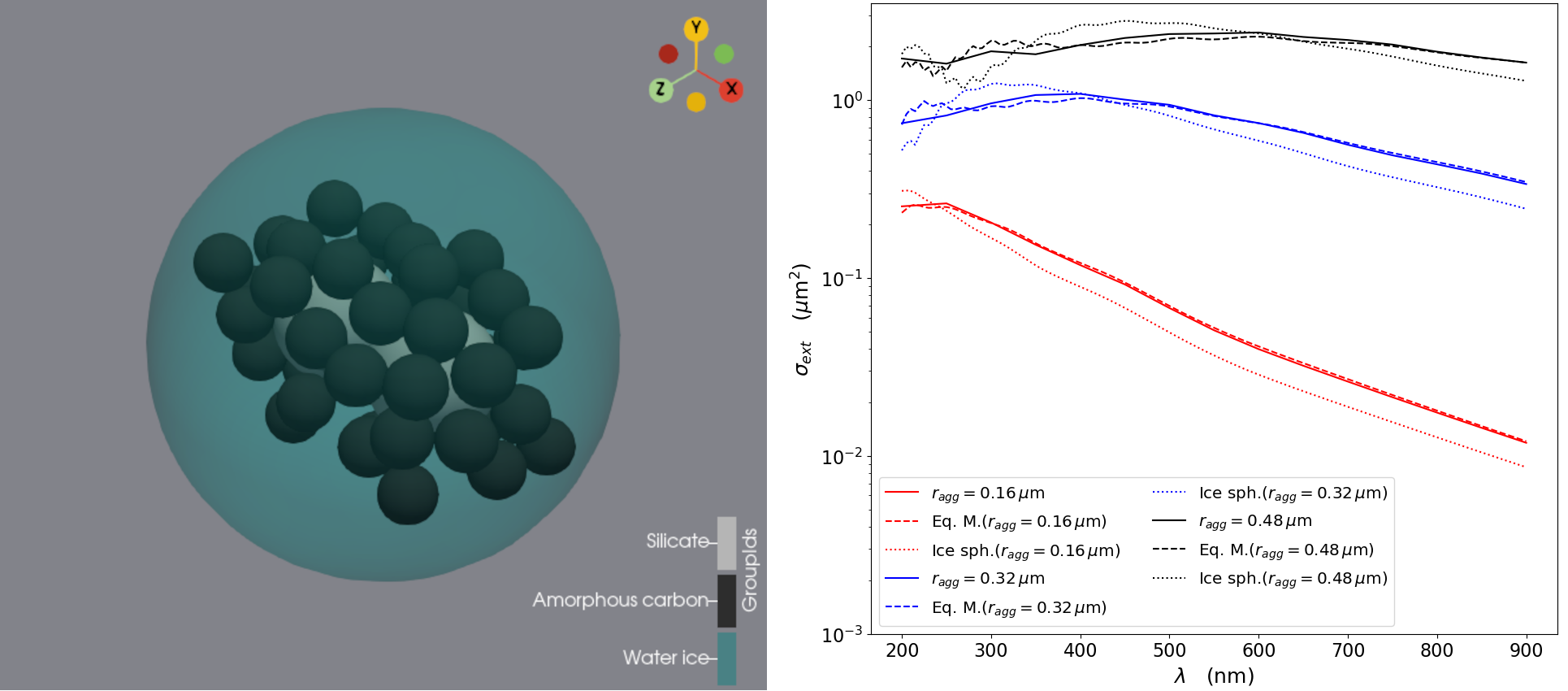}
  \end{center}
  \caption{Same as Fig.~\ref{fig_mod02}, but for the model particle embedded in an outer spherical shell of water ice. Here the predicted extinction cross-sections for the model particles (continuous lines) are compared with the corresponding equivalent mass spheres (dashed lines) and with pure ice spheres having the same radii (dotted lines). \label{fig_mod04}}
\end{figure*}
In order to thoroughly explore the matter, we need the ability to test for the effects of grain composition, size, and structural properties, ideally exploring various types of particle constituents. To illustrate the possibilities offered by \nptm, we calculated the average extinction properties of two variants of a core-shell particle model, using an elongated particle composed by a silicate core and covered by a layer of amorphous carbon and an equal composite particle embedded in a spherical outer shell of water ice. The core of our dust grain model is represented by $2$ silicate spheres, surrounded by a mantle of $44$ amorphous carbon spheres. The silicate spheres have $2.25$ times the radius of the carbon ones and the whole structure is scaled in such a way that the entire particle fits within a maximum radius of $r_{agg} = 0.158\, \mu$m, $0.316\, \mu$m, and $0.474\, \mu$m. We assumed that the optical properties of astronomical silicates correspond to the ones described by \citet{Draine84}, that those of amorphous carbon are such as given by \citet{Palik91}, and those of water ice are the ones reported by \citet{Warren08}. These dielectric functions are remarkably smooth in the spectral interval considered for all the materials, almost entirely real and with the largest variation concentrated in the UV side of the spectrum (the dielectric functions of all the materials considered in this work are shown in appendix A). The results of the multi-sphere T-matrix implementation for the particle with the smallest scale were compared with DDA-based average extinction cross-sections for the same particle geometry, resulting in an excellent agreement among the two methods. The role of particle structure at different scales, on the contrary, is tested by comparing the T-matrix calculation over the wavelength range running between $0.2\, \mu \mathrm{m} \leq \lambda \leq 0.9\, \mu \mathrm{m}$ with the corresponding equivalent mass spherical particles, solved within the Mie theory approach.

Looking at the results obtained for the direction averaged extinction curve of the ``naked'' Si-C composite particle, plotted in Fig.~\ref{fig_mod02}, we observe a growing deviation from the predictions achieved in pure spherical symmetry as the particle size is increased, particularly in the short wavelength regime. The structural effects appear to become less important for small particles, or in presence of an outer spherical coating, which mitigates the differences to the extent that the elongated particle with a spherical ice mantle becomes practically coincident with the perfectly spherical case in the small size regime, though still different from a perfectly spherical pure ice particle (Fig.~\ref{fig_mod04}).

\subsection{Particles in exoplanetary atmospheres}
\label{sec-exoatmo}

The study of exoplanetary systems has witnessed an outstanding advance through the development of high resolution spectroscopy capable instrumentation. The data produced by modern observing facilities, such as JWST and VLT, not only allowed for systematic detections of molecular compounds in the atmospheres of planets outside the Solar System, but also provided evidence for seasonal changes, weather-related features, and altimetry profiles \citep{Mukherjee26}. All such details have a huge impact on the characterization of the atmosphere (e.g. through the investigation of bio-signatures, layering, etc.), as well as on unraveling the still poorly-understood process of planetary formation and evolution. Detailed observations of planetary body atmospheres demonstrate that the formation of aerosols, in the form of clouds and hazes, have major effects on the optical properties of the observed systems, often adding more or less strong gray absorption contributions \citep{Kreidberg14, Sing16}. The typical particles that can form are controlled by the physical conditions of the host planet and they can cover a wide range of particle sizes, going from sub-micrometer scales, up to several micrometers ($r_{agg} \simeq 10\, \mu$m).

\begin{figure}
  \begin{center}
    \includegraphics[width=0.48\textwidth]{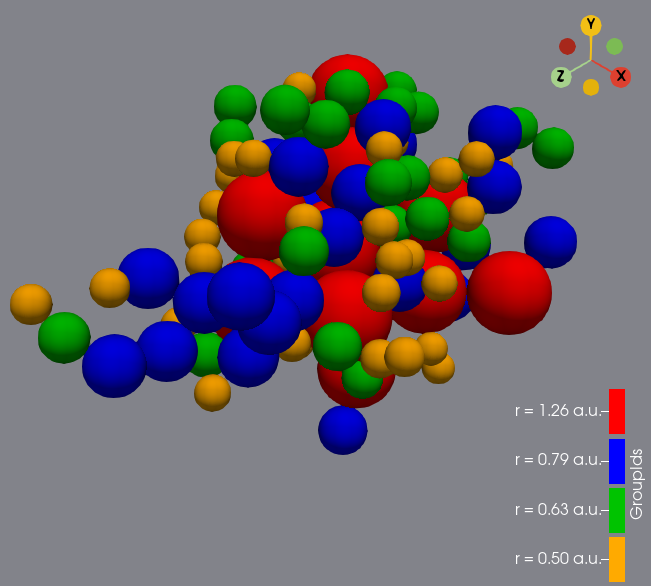}
  \end{center}
  \caption{Example of a random aggregate obtained as the combination of $100$ spherical monomers, belonging to $4$ distinct component types, each one characterized by its own radius. Different colors are used to show the different component groups. \label{fig_mod03_3d}}
\end{figure}
To illustrate the possible effects of different particle structure and composition, we constructed a random aggregate of $100$ spheres, belonging to $4$ different classes, each characterized by a specific radius, combining them with a simple ballistic growth algorithm, as shown in Fig.~\ref{fig_mod03_3d}. We evaluated the average extinction footprint of this type of particle in the IR wavelength range spanning $1\, \mu \mathrm{m} \leq \lambda \leq 25\, \mu \mathrm{m}$, characteristic of the best available (and expected) high resolution spectroscopy observations. We scaled the particle size to let it fit within a radius of $r_{agg} = 1\, \mu$m, $2\, \mu$m, $3\, \mu$m, and $5\, \mu$m. We tested our model with various compositions, using materials such as \textit{forsterite} ($\mathrm{Mg Si O_4}$) and \textit{spinel} ($\mathrm{Mg Al_2 O_4}$), which are able to condensate in the atmospheres of hot-Jupyter planets, again repeating the calculation of the smallest scale particle with DDA, as well. To perform these calculations, we used the dielectric functions provided by \citet{Jaeger03}, for the case of forsterite, and by \citet{Fabian01} for that of spinel.

\begin{figure*}
  \begin{center}
    \includegraphics[width=0.95\textwidth]{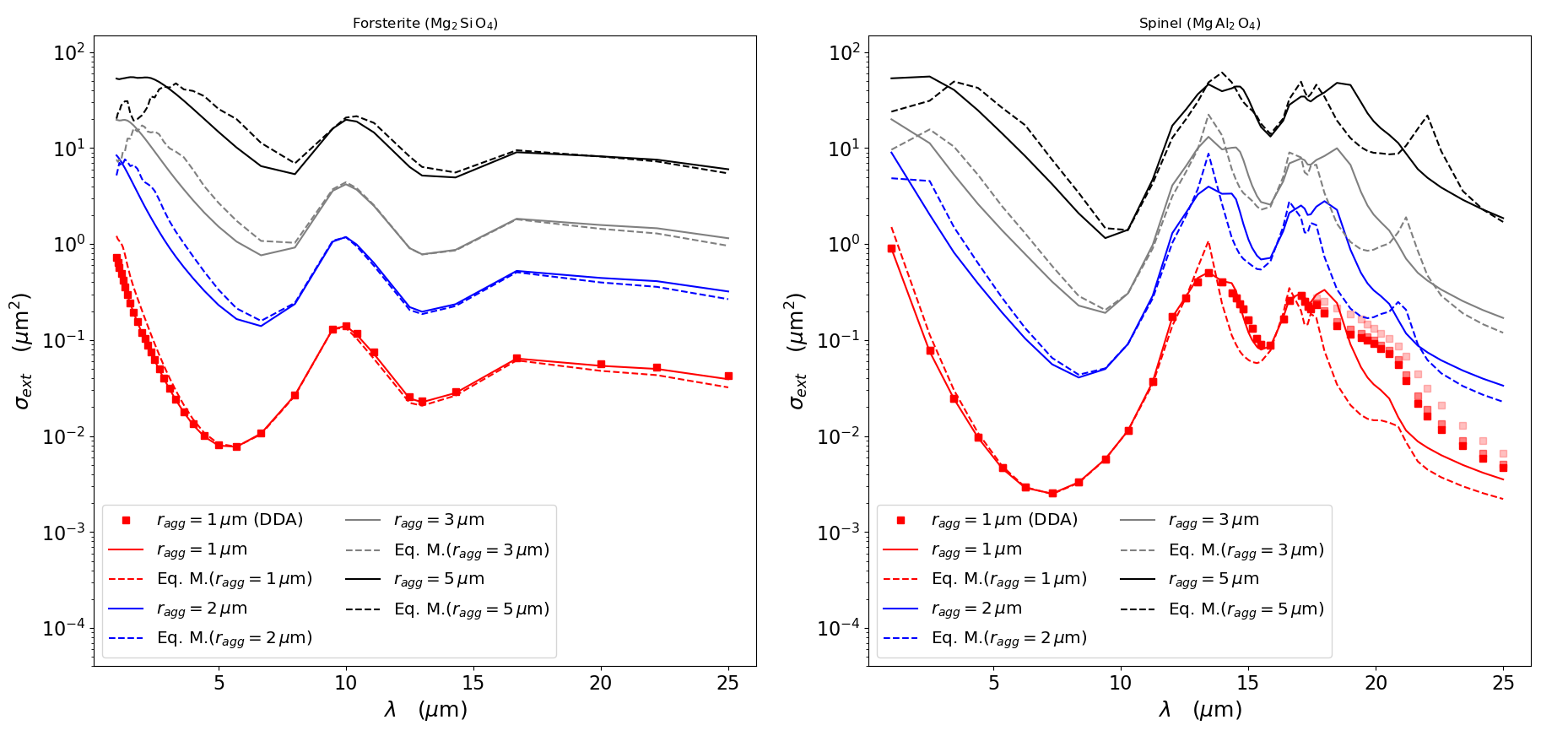}
  \end{center}
  \caption{Average extinction cross-sections, computed in the wavelength range $1\, \mu \mathrm{m} \leq \lambda \leq 25\, \mu \mathrm{m}$ for a \textit{forsterite} (left panel) and a \textit{spinel} (right panel) scaled version of a model particle, based on the structure shown in Fig.~\ref{fig_mod03_3d}, that fits within a maximum radius of $r_{agg} = 1\, \mu$m (red continuous lines), $2\, \mu$m (blue continuous lines),  $3\, \mu$m (gray continuous lines), and $5 \mu$m (black continuous line). All the particle models are compared with the corresponding Mie solutions for spherical particles of equivalent mass (dashed lines), while the small scale models are also compared with averaged DDA calculations (red squares). For the spinel Reststrahlen band, we show the results of DDA calculations for dipole sizes of $d = 15\,$nm (computed), $7.5\,$nm (computed), and $1.7\,$nm (extrapolated), using markers with increasing opacity. \label{fig_mod03}}
\end{figure*}
The results of this set of models are shown in Fig.~\ref{fig_mod03}. It can be again appreciated that we expect an increasing importance of particle structure for bigger particles, with respect to smaller ones and that the effects are, at a first glance, more marked in the short wavelength regime, but with some notable exceptions. In first place, indeed, we note that the pronounced extinction features seen around $9.7\, \mu$m and $18\, \mu$m in the case of forsterite \citep[which are well known properties of silicate materials, see e.g.][]{Gillett75, Chiar06} tend to be independent of the structural effects explored by our example models. This effect appears to arise equally in our T-matrix calculation and in the corresponding DDA simulation. Conversely, the extinction features that characterize the case of spinel exhibit rather strong effects which may, in some cases, affect the model extinction cross-section, leading both the T-matrix and the DDA prediction to deviate from the spherical case by factors as high as $10$. Upon detailed investigation, we note that the silicate features of forsterite occur in a spectral region where the extinction cross-section is dominated by the absorption component, which appear to be less sensitive to the detailed structure of the particle model. The case of spinel, on the contrary, is characterized by abrupt changes in the dielectric functions, which render the predicted cross-sections more strongly dependent on geometry. In similar cases, it is known that substantial increase in the resolution of the model grid is needed to obtain convergence with DDA \citep{Yurkin10, Yurkin11}.

\begin{figure*}
  \begin{center}
    \includegraphics[width=0.95\textwidth]{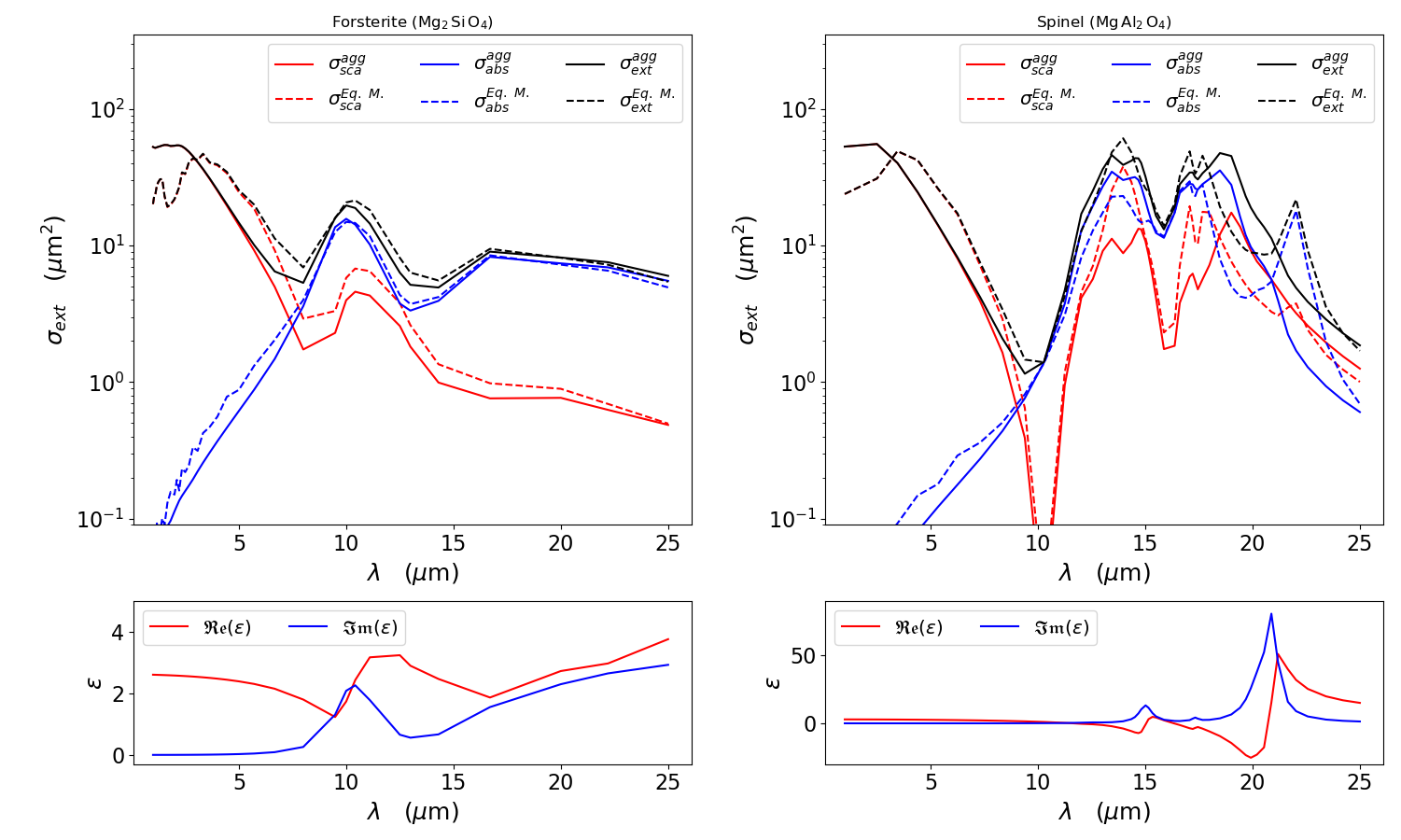}
  \end{center}
  \caption{UPPER ROW: forsterite (left) and spinel (right) extinction cross-sections (black lines), obtained as combination of scattering (red lines) and absorption components (blue lines), for the case of the $r_{agg} = 5\, \mu$m particle. We use continuous strokes for the model aggregates and dashed lines for the Mie solution of the corresponding equivalent mass spheres. LOWER ROW: forsterite (left) and spinel (right) dielectric functions. We plot as red lines the real part of $\varepsilon$ and as blue lines the corresponding imaginary parts. We note that the sharp variations of the spinel dielectric function around $15\, \mu$m and $21\, \mu$m lead to extinction features predicted by the spherical approximation which are smoothed away in the aggregate case. \label{fig_components}}
\end{figure*}

We further explored the matter in Fig.~\ref{fig_components}, where we plot the extinction cross-sections of the $5\, \mu$m size particles of forsterite and spinel, as a combination of their scattering and absorption contributions, together with the real and imaginary parts of their dielectric functions. As summarized by Fig.~\ref{fig_components}, the results obtained for MgAl$_2$O$_4$ spinel show a markedly different behavior from that of forsterite and appear to be strongly influenced by the peculiar infrared dielectric response of spinel. The spectral region where the real part of the dielectric function becomes negative is consistent with the typical behavior of a Reststrahlen material, such as spinel. In this range, the response is governed by infrared-active lattice vibrations rather than by free-electron plasmons. Nevertheless, from an electromagnetic point of view, the material can sustain phonon-polaritonic, plasmon-like resonances. As far as $\mathfrak{Re}(\varepsilon)$ approaches negative values, the polarizability of the individual monomers increases significantly, enhancing both local-field effects and multi-polar coupling. This leads to a pronounced scattering contribution, much larger than in the forsterite case.
Moreover, the comparison between the aggregate and the equivalent-mass sphere suggests a clear and strong aggregation effect. In particular, the absorption peak of the equivalent sphere around $22\, \mu$m becomes just a shoulder when we deal with the aggregate. This can be interpreted as the result of electromagnetic coupling among the monomers in the aggregate, which broadens, shifts and redistributes the resonance over several collective modes. Therefore, this feature should not be interpreted as a new intrinsic resonance of spinel, but rather as a morphology-induced modification of the polaritonic response, similarly to the effects predicted for metallic nano-particles \citep{Sharma26}.

The comparison with forsterite is instructive. Forsterite shows a more conventional and expected behavior: absorption dominates, scattering remains comparatively weaker, and the aggregate response follows the equivalent-sphere approximation more closely. By contrast, spinel exhibits a much stronger sensitivity to aggregation because its dielectric function enters a Reststrahlen-like regime, where small changes in morphology and local electromagnetic coupling can strongly affect the extinction, absorption and scattering spectra. The implications of our models suggest that, with the assumed dielectric functions, the extinction features dominated by absorption and characterized by regular dielectric functions are weakly dependent on particle structure and more reliable indicators of quantitative effects than features entering the Reststrahlen domain.

Clearly a more extensive investigation of the impact of particle structure, size, and composition requires exploring a larger parameter space in terms of possible aggregates, than the one explored in this study. A project aiming at the systematic calculation of the effects of particle structure and composition with \nptm\ produced the data collected by \citet{aresu_db_01, aresu_db_02, aresu_db_03}, while a thorough description of the subsequent implications is given by Aresu et al. (in prep.).

\section{Computing performance and requirements}
\label{sec-perf}

The motivation to develop \nptm\ is to provide a fast and flexible implementation of the multi-sphere T-matrix formalism, able to work with particle models extending over a wider range of size parameters than what was previously possible, provided that the hardware resources support the calculation. While the code has been designed to take advantage from high performance computing architectures (HPC), it is also able to run on commercial hardware, including laptop and workstations, since its advanced functions, such as offloaded GPU calculations and parallel execution, are only enabled as optional features, depending on the support offered by the host system. The compatibility of the code is tested against different \textit{C++} compilers, including both open source versions, such as \textit{GNU} \texttt{g++} (using compiler version \texttt{v11} or higher) and \texttt{CLANG} (\texttt{v16} or higher), as well as \textit{Intel}'s \texttt{icpx} (\texttt{v2025.1.0}). The consistency of \nptm\ calculation results with the legacy implementation of Borghese, Denti and Saija (written in \texttt{FORTRAN66} and also distributed along with \nptm) is verified through a set of test models, included in the code's distribution.

Not all the models considered in this study could be managed by the original serial implementation, but we further tested the consistency of the results between \nptm\ and the original implementation for the naked Si-C particle model of Fig.\ref{fig_mod02}, running the serial code, compiled with \textit{GNU} \texttt{gfortran v15.2.0} and fast optimization (\texttt{-O3}), for the particle models scaled to fit in $r_{agg} = 0.158\, \mu$m and $r_{agg} = 0.316\, \mu$m. All the numerical results were found in agreement with the serial implementation, within relative differences better than $0.1\%$. In addition, by running the calculations on an \textit{ASUS Zenbook} with $32\,$GiB of RAM, a $20$-thread capable $13^\mathrm{th}$ Gen Intel$\circledR$ Core$^\mathrm{TM}$ i9-13900H CPU, and a NVIDIA$\circledR$ GeForce RTX$^\mathrm{TM}$ 4060 Laptop GPU with $8\,$GiB VRAM, we observed that the serial implementation needed $1651.42\,$s to solve the $r_{agg} = 0.158\, \mu$m model case and $34705.43\,$s to solve the $r_{agg} = 0.316\, \mu$m model case for $15$ different wavelengths, whereas the corresponding computing times of \nptm, using the \texttt{OpenMP} thread hierarchy setting \texttt{OMP\_NUM\_THREADS=5,4}, took respectively only $49.96\,$s (with a speed-up factor of $33.04$) and $398.40\,$s (speed-up factor of $87.11$), showing extremely promising weak scaling capabilities.

With the increasing complexity of models, the most stringent limitation of \nptm\ is its memory consumption, since the T-matrix of particles composed by tens to hundreds of spherical units and size parameters of the order of $10$ can require gigabytes of memory to be handled. If using a GPU, the T-matrix needs to fit in the available device RAM, to perform in-place inversion.\footnote{The device needs to store at least $3$ copies of the matrix, if using the optional feature of iterative inversion refinement.} In addition, heavy optimization of matrix inversion operations, together with the necessity to store radiation field expansions, introduce further overheads, which are implementation dependent, but have been observed to reach up to approximately $10$ times the size of the T-matrix, for every radiation wavelength, using the most aggressive speed optimizations. As a consequence, the calculation of larger particle models, such as those designed to represent exoplanetary aerosols, had either to be executed with less aggressive \texttt{OpenMP} parallelization or to be submitted as batch jobs to computing systems featuring more powerful hardware. The batch job choice was almost mandatory to solve the $r_{agg} = 5\, \mu$m, since the corresponding T-matrices had sizes up to $3.81\,$GiB. These calculations were carried out by CPU partition nodes of the computing farm hosted by the INAF - Astronomical Observatory of Cagliari (8 nodes equipped with $32$ $2 \times$ AMD Epyc 7313 CPUs and $503\,$GiB RAM). Running the same forsterite case on $37$ wavelengths using a single node and an \texttt{OpenMP} setting \texttt{OMP\_NUM\_THREADS=4,12} (48 threads working scattered over 4 simultaneous wavelengths) led to obtain the solution in $6026.80\,$s, while executing the same calculation on $2$ nodes and an \texttt{OpenMP} setting \texttt{OMP\_NUM\_THREADS=8,8} (128 threads working on 8 simultaneous wavelengths) achieved completion in $2305.17\,$s, suggesting that the code is able to achieve a nearly linear strong scaling.

Due to time and energy saving reasons, there was no possibility to benchmark the code speed-up factors on the most computationally intensive calculations, because they are not easily solved by equivalent alternative approaches. To give an idea of the possibilities offered by \nptm, the \texttt{ADDA} validation test for the $r_{agg} = 0.158\, \mu$m ISM particle took $1248.48\,$s using $8$ MPI processes, against $49.96\,$s taken by $20$ \texttt{OpenMP} \nptm\ threads working on $5$ simultaneous wavelengths. Even more dramatic differences were observed for the exoplanetary particle models. We did not compute the resulting speed-up factors for these cases, because finding the best configuration to define equivalent particle models for \texttt{ADDA} is not within the scope of the present work. We noted however, that \texttt{ADDA} was able to obtain equivalent results using from $10$ to $30$ times less RAM resources than \nptm.

\section{Conclusions}
\label{sec-discussion}

In this work we describe the application of \nptm, a new high performance implementation of the multi-sphere T-matrix formalism, to a set of illustrative particle models aiming at the investigation of problems such as material mixtures and structural particle complexity for interstellar dust grains and exoplanetary atmospheric aerosols. A distinguishing feature of this method is its inherent capability to describe particle components characterized by different size, material composition and layer structure. The calculations executed with \nptm\ proved to be faster than similar calculations implemented in a serial T-matrix formalism, by factors of more than $30$ and $80$ times per wavelength for the interstellar dust grain model of growing size, presented in Section~\ref{sec-dust}. They also extend the domain of T-matrix solutions to the scale of large condensed particles, which could not be investigated previously within reasonable times. Therefore, the development of \nptm\ represents a significant advancement in the accessibility and execution of T-matrix based calculations. With respect to the benchmark study presented in \citet{LaMura26}, the applications described in this work show that \nptm\ acquired more flexibility, particularly with the conversion of features such as the matrix inversion algorithms and the GPU offload capabilities from compilation settings to runtime options and with the inclusion of auxiliary scripts, helping with the definition of input model sets and the manipulation of output results.

We explored the scientific relevance of the new code features by performing an illustrative investigation of the effects of particle size, geometry and composition on the spectral footprint that various types of particles are expected to leave in wavelength dependent extinction curves and transmission spectra. Although the variety of considered model is still limited, we can already point out some suggestive effects. If, on the one hand, we see that particle structural properties can have a substantial effect on the extinction curve, such as smoothing out sharp features connected with the material optical properties, thus leaving less evident footprints in irregular aggregates, with respect to perfectly spherical particles, on the other, strong absorption bands may be more stable and, therefore, provide more reliable handles to perform quantitative spectroscopic analysis. In any case, these results imply that caution is necessary when using strong material features to assess the absence of presence, and infer the amount, of specific components: the direct, naive comparison of astronomical observations of extinction with bulk absorption cross sections of solid materials, without considering the morphology of the particles, may be very misleading.

With the publication of the first high resolution JWST observations of transiting exoplanets, and the perspective of collecting thousands of spectra from the upcoming ARIEL mission, we expect that the \nptm\ suite will become a strategic asset. It will provide a fundamental contribution to the modeling of radiation scattering and extinction, offering an unprecedented degree of accuracy for astrophysical investigations concerning the properties of the interstellar medium, the propagation of radiation in exoplanetary atmospheres, and, more generally, affecting all the applications where a detailed treatment of the radiation transfer is needed.

\begin{figure*}
  \begin{center}
    \includegraphics[width=0.95\textwidth]{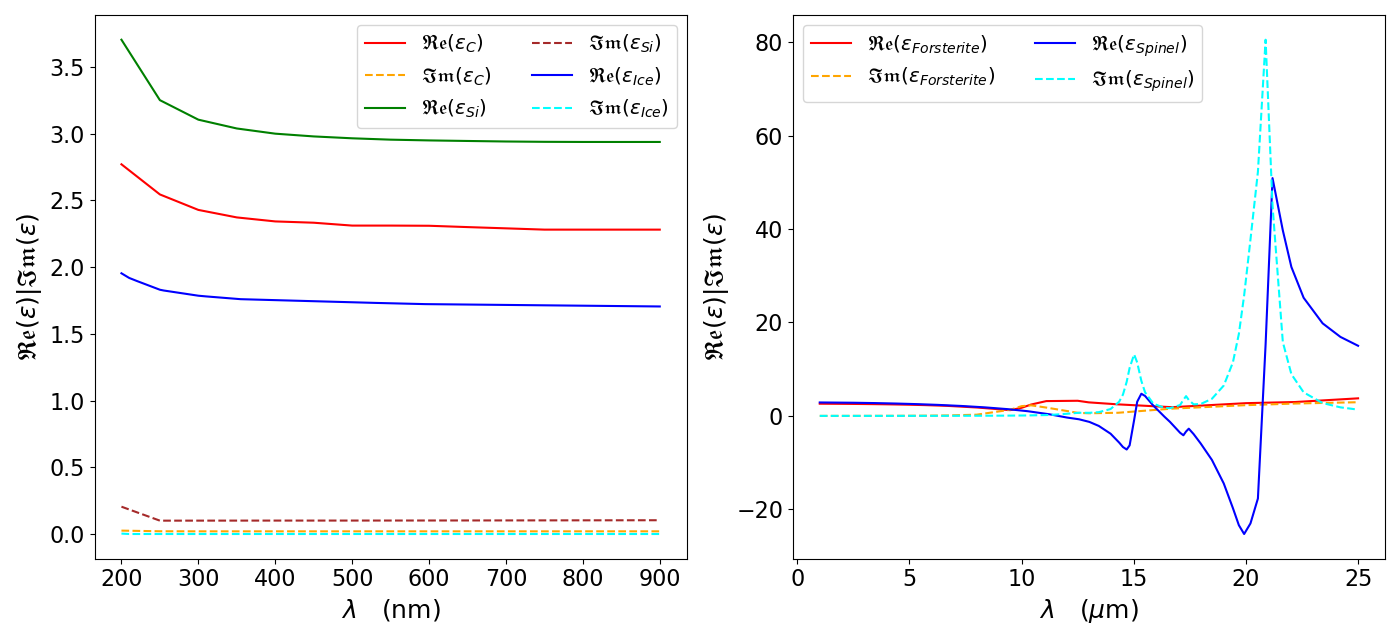}
  \end{center}
  \caption{Illustration of the dielectric functions for the materials considered in this study. Different materials are plotted with different colors, while continuous and dashed lines are used to represent, respectively, the real and the imaginary parts. \label{fig_diel_funcs}}
\end{figure*}
\section*{Acknowledgments}

Supported by Italian Research Center on High Performance Computing Big Data and Quantum Computing (ICSC), project funded by European Union - NextGenerationEU - and National Recovery and Resilience Plan (NRRP) - Mission 4 Component 2 within the activities of Spoke 3 (Astrophysics and Cosmos Observations) and by the Italian Ministry of University and Research (MUR), through PRIN Exo-CASH (PRIN 2022 project no. 2022J7ZFRA). This work was completed in part at the CINECA GPU HACKATHON 2024, part of the Open Hackathons program. The authors would like to acknowledge OpenACC-Standard.org for their support. The \texttt{NP\_TMcode} software is the parallel implementation of the T-matrix formalism, developed and maintained at INAF - Astronomical Observatory of Cagliari as a contribution to the activities of ICSC \textit{Spoke 3} ``Astrophysics and Cosmos Observations'', on the basis of the codes originally written by F. Borghese, P. Denti and R. Saija. It is registered in the \textit{Astrophysics Source Code Library} at \href{https://ascl.net/2510.003}{https://ascl.net/2510.003}.

\section*{Data availability}
All the input data and the results of the models presented in this work are distributed via \textit{Zenodo} at \url{https://doi.org/10.5281/zenodo.22675006}.

\printcredits

\begin{table*}
\caption{Summary of the parameters used to define DDA model grids to obtain equivalent geometries for comparison with the corresponding T-matrix calculations. The table reports, in order, the model name, the largest particle span along the Cartesian axes of the grid ($X$, $Y$, or $Z$), the number of cells filled by the model particle, the size of each dipole, the shape of the bounding box of the particle, the DDA solver (either LDR for lattice dispersion relation, or FCD for filtered coupled dipoles),  and the shortest wavelength covered by the calculation. \label{tab-dda}}
\begin{tabular*}{\tblwidth}{@{}LCCCCCC@{}}
\toprule
 Model & Size & $n_{dip}$ & $d$ & Grid shape & Solver & $\lambda_{min}$ \\
\midrule
Naked Si-C & $0.320\, \mu$m & $34958$ & $4.0\,$nm & $80 \times 40 \times 40$ & LDR & $0.2\, \mu$m \\
Exoplanetary (forsterite) &  $1.710\, \mu$m & $59281$ & $15.0\,$nm & $68 \times 87 \times 114$ & LDR & $1.0\, \mu$m \\
Exoplanetary (spinel, coarse) &  $1.710\, \mu$m & $59281$ & $15.0\,$nm & $68 \times 87 \times 114$ & FCD & $1.0\, \mu$m \\
Exoplanetary (spinel, fine) &  $1.710\, \mu$m & $474612$ & $7.5\,$nm & $136 \times 175 \times 228$ & FCD & $1.0\, \mu$m \\
\bottomrule
\end{tabular*}
\end{table*}
\section*{Appendix A - Dielectric functions}

The optical properties of the materials considered in this study are defined through their complex dielectric functions. In this study, we used the dielectric functions of amorphous carbon \citep{Palik91}, astronomical silicates \citep{Draine84}, water ice \citep{Warren08}, forsterite \citep{Jaeger03}, and spinel \citep{Fabian01}. Plots of the real and imaginary parts of these functions are shown in Fig.~\ref{fig_diel_funcs}.

\section*{Appendix B - DDA model implementation}

The average extinction cross-sections for the small scale particle models presented in this study were computed using both the T-matrix and the DDA formalism, to verify the consistency of these two numeric approaches. In order to produce comparable simulations, we converted the multi-sphere aggregate particles that define the input of the T-matrix calculation into grids of fixed volume discrete cubic cells, each occupied by dipoles with polarizability determined by the complex refractive index of the material located in the cell position. The recommended grid spacing to apply DDA calculations is \citep{Yurkin11}:
\begin{equation}
    d = \dfrac{\lambda}{10 |m|}, \label{eq_rec_dda}
\end{equation}
where $m$ is the complex refractive index. Eq.~(\ref{eq_rec_dda}) holds as a reference for materials with $|m - 1| < 2$, while larger refractive index require much finer grid definitions. This last condition is marginally violated by astronomical silicates for $\lambda < 300\,$nm (where $|m - 1| = 2.708$), as well as by forsterite in the long wavelength regime ($|m - 1| = 3.758$ at $\lambda = 25.0\, \mu$m), and strongly violated by spinel in the Reststrahlen band ($|m - 1| = 12.193$ at $\lambda = 15.0\, \mu$m and $|m - 1| = 80.957$ at $\lambda = 20.9\, \mu$m). The details of the parameters used to convert our T-matrix models into the corresponding input for DDA are summarized in Table~\ref{tab-dda}.

As discussed in Section~\ref{sec-exoatmo}, the grid spacing chosen for our models is generally able to satisfy the conditions posed by Eq.~(\ref{eq_rec_dda}), except for the plasmonic resonance of spinel, which would require a grid resolution of the order of $1\,$nm. Since we can not presently access a system with the necessary hardware resources to solve such a model, we evaluated the extinction cross-sections as a function of wavelength using two different grid spacing (a coarse grid made of dipoles with $d = 15.0\,$nm and a fine grid with $d = 7.5\,$nm), then we performed an extrapolation, assuming that the result for this case approaches convergence as a power-law function of the dipole size:
\begin{equation}
    \sigma_{ext}^{(d)} (\lambda) = \sigma_{ext}^{(7.5\, \mathrm{nm})} (\lambda) \left( \dfrac{d}{7.5\, \mathrm{nm}} \right)^{-\alpha(\lambda)}
\end{equation}
where $\alpha(\lambda)$ was obtained from the comparison of the coarse grid with the refined one and the extrapolation was evaluated for $d = 1.7\,$nm.

\end{document}